\documentclass{aa}
\usepackage{graphicx}
\usepackage{txfonts}
\usepackage{natbib}
\bibpunct{(}{)}{;}{a}{}{,} 

\begin{document} 

\title{2MASS Galaxies in the Fornax Cluster Spectroscopic Survey\thanks{
Our matched catalogues are available in electronic form
at the CDS via anonymous ftp to cdsarc.u-strasbg.fr (130.79.128.5)
or via http://cdsweb.u-strasbg.fr/cgi-bin/qcat?J/A+A/} }

\author{R.A.H. Morris \inst{1} \and S. Phillipps \inst{1} \and
J.B. Jones \inst{2} \and M.J. Drinkwater \inst{3} \and M.D. Gregg
\inst{4,5} \and W.J. Couch \inst{6} \and Q.A. Parker \inst{7,8} 
\and R.M. Smith \inst{9} }

\institute{Astrophysics Group, Department of Physics, University of
Bristol, Bristol BS8 1TL
\and
Astronomy Unit, School of Mathematical Sciences, Queen Mary
University of London, Mile End Road, London E1 4NS
\and
Department of Physics, University of Queensland, QLD 4072, Australia
\and
Department of Physics, University of California Davis, CA
95616, USA
\and
Institute for Geophysics and Planetary Physics, Lawrence
Livermore National Laboratory, Livermore, CA 94550, USA
\and
School of Physics, University of New South Wales, Sydney
2052, Australia
\and
Department of Physics, Macquarie University, Sydney, NSW,
Australia
\and
Anglo-Australian Observatory, Epping, NSW 1710, Australia
\and
School of Physics and Astronomy, Cardiff University, Cardiff, CF24 3AA, UK
}


\date{Received 1 July 2005 /
Accepted 27 August 2007)}
 


\abstract 
{The Fornax Cluster Spectroscopic Survey (FCSS) is an all-object
survey of a region around the Fornax Cluster of galaxies undertaken
using the 2dF multi-object spectrograph on the Anglo-Australian
Telescope. Its aim was to obtain spectra for a complete sample of all
objects with $16.5<b_j<19.7$ irrespective of their morphology
(i.e. including `stars', `galaxies' and `merged' images).} 
{We explore the extent to which (nearby) cluster galaxies are present
in 2MASS. We consider the reasons for the omission of 2MASS galaxies
from the FCSS and vice versa.} 
{We consider the intersection (2.9 square degrees on the sky) of our
data set with the infra-red 2 Micron All-Sky Survey (2MASS), using
both the 2MASS Extended Source Catalogue (XSC) and the Point Source
Catalogue (PSC). We match all the XSC objects to FCSS counterparts by
position and also extract a sample of galaxies, selected by their
FCSS redshifts, from the PSC.}
{We confirm that all 114 XSC objects in the overlap sample are
galaxies, on the basis of their FCSS velocities. A total of 23 Fornax
Cluster galaxies appear in the matched data, while, as expected, the
remainder of the sample lie at redshifts out to $z = 0.2$ (the spectra
show that 61\% are early type galaxies, 18\% are intermediate types
and 21\% are strongly star forming).The PSC sample turns out to
contain twice as many galaxies as does the XSC. However, only one of
these 225 galaxies is a (dwarf) cluster member. On the other hand,
galaxies which are unresolved in the 2MASS data (though almost all are
resolved in the optical) amount to 71\% of the non-cluster galaxies
with 2MASS detections and have redshifts out to $z=0.32$.} {}



\maketitle

\titlerunning{2MASS Galaxies in the Fornax Cluster Spectroscopic Survey}
\authorrunning{R.Morris et al.}


\section{Introduction}

There are obvious differences between infra-red and optically selected
catalogues of galaxies. In this note we discuss the results obtained
by cross-referencing objects in our `all-object' Fornax Cluster
Spectroscopic Survey \citep[FCSS:][]{drinkwater00a} with those in the
infra-red 2 Micron All-Sky Survey \citep[2MASS:][]{jarrett00} in the direction of the Fornax cluster. 
Brief details of these surveys are presented in Section 2.  At first
glance, these two catalogues are poorly matched, in that 2MASS is a
much larger area but comparatively shallow survey, relative to the
FCSS.  However, the FCSS is designed to cover {\em all} types of
object within its magnitude range, so a wide range of objects may
inhabit the overlap in parameter space. Many of the objects in common
are stars, but we will concentrate here on the galaxy component, and
especially on the cluster members. \citet{cole01}, henceforth C2001,
have previously matched 2MASS to the general 2dF Galaxy Redshift
Survey \citep[2dFGRS:][]{colless01}. However, here we consider the 2MASS XSC objects
which would not have had 2dFGRS counterparts and the galaxy content of
the unresolved sources in the PSC.

\section{The Catalogues}

The 2dF multi-fibre spectrograph
\citep{lewis02}
on the Anglo-Australian Telescope, provides the opportunity
to make truly complete spectroscopic surveys of a given area on the
sky, down to well determined, faint limits, irrespective of target type.
The FCSS exploited this by fully surveying a region centred on the Fornax
Cluster of galaxies \citep{drinkwater00a,drinkwater00b}.

In common with the main 2dFGRS (which did not cover the Fornax area), 
our input catalogue for the FCSS comes from
UKST Sky Survey plates digitised by the APM machine \citep{irwin94},
but unlike other surveys, we avoid any morphological pre-selection and
include {\em all} objects, both resolved and unresolved (i.e.\ `stars'
and `galaxies') between our sample limits at $16.5 \leq b_{j} \leq
19.7$, as well as the confused sources classified as `merged'. 
We report results here from the central 2 degree diameter field from
the FCSS (centred close to the first ranked cluster galaxy NGC~1399)
for which the redshift measurements are 92\% complete \citep{deady02}. In order to obtain a greater overlap with 2MASS, 
we have
also added in the galaxies brighter than $b_j = 16.5$ in our survey
area by utilising our companion survey \citep{drinkwater01a} which
used the FLAIR spectrograph on the UK Schmidt Telescope, as well as
literature sources. Note that these were morphologically selected as
galaxies but we do not expect any confusion of compact galaxies with
stars at these very bright magnitudes.

The 2MASS\footnote{2MASS is a joint project of the University of
Massachusetts and the Infrared Processing and Analysis Center/
California Institute of Technology. Funding for the survey has been
provided by NASA and the NSF} Second Incremental Data Release (March
2000) (also used by C2001) includes both a Point Source Catalogue
(PSC) and an Extended Source Catalogue (XSC). The XSC should be be
complete for all galaxies brighter than $K_s = 13.5$ and contains
objects down to $K_s \simeq 14.4$, depending on their degree of
extension, surface brightness etc.  Given the resolution of 2MASS, XSC
objects are typically in excess of $10''$ in diameter
\citep{jarrett00}.  For the photometry below we used the extrapolated
`total' magnitudes.

The FCSS area was used to define our potential sky coverage, viz. a
$2^{\rm o} $ diameter circle centred at $03^{\rm h} 38^{\rm m} 29^{\rm
s}$, $-35^{\rm o} 27'$ (J2000).  We used the `Gator' query tool in the
IPAC IRSA web interface to extract sources from the 2MASS PSC and XSC
satisfying the same area constraint. In fact, Fornax lies close to the
edge of the region surveyed for the 2MASS Second Incremental Data
Release, so the overlap region is slightly less than the nominal 2dF
area and covers 2.9 square degrees, containing 3204 FCSS objects, of
which the great majority are Galactic stars or galaxies behind the
cluster. In reprocessing data for the later All Sky Data Release, several
former PSC objects were reclassified into the XSC. In what follows we
have used this updated information but retained our original field
area limit. \footnote{Strictly speaking, 6 objects were reclassified
{\em out} of the XSC in the all sky 2MASS catalogue, but as we confirm
that they really are galaxies, we have retained them in the `extended'
sample. They make no significant difference to any of the statistics
presented later.}

A generous search radius of $3''$ was used in matching 2MASS to FCSS
objects although it is already known that for point sources 2MASS
positions agree with those from the APM catalogue to $\sim 0.5''$.
The relatively large search radius of $3"$ was retained to allow
matching of large, possibly irregular, galaxies with larger centroid
uncertainties. A significant contribution to the difference in object
centroids between 2MASS and APM will be variations in colour for
large irregular galaxies. In total 114 XSC and 2021 PSC objects (not
also in the XSC) match with ones in the FCSS or the brighter
sample. Our matched catalogues are available from the CDS in electronic form.

\section{2MASS XSC Objects}

Of the 114 2MASS XSC objects in the area, 84 are matched in 
the FCSS itself, 28 are in the brighter
FLAIR sample and two are 15th magnitude galaxies in the
\citet{ferguson89} Fornax Cluster 
Catalogue (FCC).  All are confirmed as galaxies, with
redshifts showing them to be in or behind the Fornax Cluster.
This compares with $\sim$ 11\% of XSC objects not matched in the
2dFGRS in C2001. In our area there are no saturated or multiple stars 
in the XSC
and none wrongly classed as stars by APM, 
while 4 were erroneously classed as extended in the original XSC --
consistent with the combined tally of $\sim 7$\% in C2001 -- but the 
latter 4 were
removed in the next release. The remaining $\sim 4$\% of unmatched 2MASS
objects in C2001 were `merged' images in the $r$ band APM
images, whereas we spectroscopically observed all of these. (In fact,
in our area we have a surprisingly large number of merged images,
amounting to $\sim 22$\% of the 2MASS sample). 
  
Given the
range of colours seen for the galaxies (up to $b_j - K_s \simeq 5.5$),
one would expect the FCSS to be deep enough in magnitude to pick up
all the genuinely extended 2MASS sources in the complete sample to
$K_s = 13.5$, as was found by C2001 for the comparably deep 2dFGRS. It
is just possible for objects in the faint tail down to $K_s
\simeq 14.4$ to be missed by the FCSS if they were also of the very
reddest colours $b_j - K_s \simeq 5.4$, but as we have seen there are
no examples of this in our sample.

A total of 23 cluster galaxies ($cz
\simeq 1400 \pm 800$~km~s$^{-1}$) appear in the matched data. All bar
one is an early type according to the types assigned in
\citet{drinkwater01a}, the FCSS or NED. Note that only 2 of the
cluster objects are from the actual 2MASS/FCSS paired data, and one of
those was only included because of an erroneous input magnitude, as it
is actually significantly brighter than the FCSS bright limit. All the
rest are from the brighter spectroscopic surveys.  Thus genuine
cluster dwarfs, as sampled by the FCSS, are virtually absent from the
2MASS XSC, despite having exactly the same $b_j$ magnitudes as detected
background galaxies. This is a result of a combination of the cluster
dwarfs' blue colours and, especially, low surface brightnesses
\citep{monnierragaigne03}.

\begin{figure}
 \resizebox{\hsize}{!}{\includegraphics{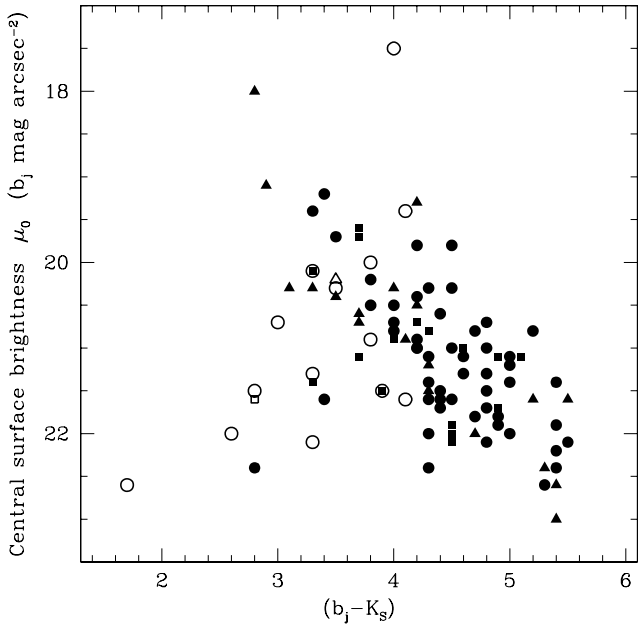}}
 \caption{Central surface brightness versus $b_j-K_S$ colour for
 matched XSC objects. Open (filled) circles are cluster (background)
early types, squares intermediate types and triangles late types.}
\label{figure1}
 \end{figure}

Matching to various template spectra, 
of the 83 FCSS galaxies (i.e. counting
the interacting pair only once), 50 (61\%) have the classic K star-like
spectrum of an early type galaxy \citep{morgan57}, 18\% 
match F or G star spectra (intermediate
type) and 21\% are emission line galaxies (late types).  
These are extremely similar to the fractions
obtained by C2001 -- 62\% early type, 22\% Sa-Sb, 16\% late type -- on 
the basis of a
completely independent spectral typing using a principal component analysis \citep[see][]{madgwick02}. The
types match reasonably well to the galaxies' $b_j - K_s$ colours,
though there exist a few late types with red colours and the cluster dwarfs 
have rather blue colours compared to \cite{poggianti97} models (see
Figure \ref{figure1}). The bluest colours are found amongst the low surface brightness
cluster members \cite[cf.][]{bell03}.\footnote{The outlier with the {\em very} blue
$b_j-K_s$, FCC222, is in the \citet{arp87} catalogue with
several classifications in different sources and
strange spectral characteristics \citep{mieske02}. Its colour may also be
rather uncertain, its $b_j$ magnitude in the SuperCOSMOS sky survey
\citep{hambly2001}
is 16.2 compared to our 15.1. \citet{mieske02} give $V=14.9$ and the eye
estimate in the FCC is $B=15.6$. SuperCOSMOS gives a
more reasonable $b_j-K_s = 2.8$.} 

There are also some
low surface brightness but very red objects, several of which are clearly edge-on spirals from their images, in
agreement with their spectral typing. Otherwise, the
overall trend for decreasing surface brightness with redder colour is
as expected from the combined effect of the k-correction reddening and
the cosmological dimming of surface brightness with redshift for the
dominant early type background galaxy population. (The redshift range
is more extended than that shown 
by C2001, as they limited their sample at $K_s = 13.2$; we can also note that
the galaxies that would have been missed by
2dFGRS selection have the same redshift range as the others).
 
\section{2MASS PSC Objects}

Now consider any `small' galaxies hidden among the 2MASS Point Source
Catalogue (PSC) objects. These may be, of course, merely distant rather
than physically small. \citet{finlator00}, in a study of stars
detected in 2MASS, have previously noted (but not discussed further)
the fact that $\approx$15\% of PSC objects correspond to resolved
images in the Sloan Digital Sky Survey (SDSS) commissioning data.

In all there are 2021 matches of 2MASS point sources with FCSS objects
out of a total of 3204 objects (both optically resolved and
unresolved) in the FCSS catalogue in the overlap region. This region
contains 3982 PSC objects which are not already covered by the
XSC. Thus only just over half (51\%) have FCSS counterparts. Given the
expected stellar number counts, a substantial number of unmatched PSC
objects will be stars brighter than $b_j= 16.5$. From the
\citet{bahcall80} models, we expect around 1200 stars brighter than
the FCSS bright limit in a 3 square degree area at $b \sim -50^{\rm
o}$. The PSC contains significantly fainter sources than does the XSC,
down to $K_s \simeq 15.5$ (with a few even fainter), so given the very
broad range of colours exhibited by both stars and galaxies \citep[][C2001]{bessell88}, many of the fainter 2MASS PSC sources
will be red enough to fall off the {\em faint} end of the FCSS
sample. Clearly anything fainter than $K_s \sim 15$ and/or redder than
$b_j-K_s \sim 4.5$ is likely to be lost.  Going the other way, a
smaller fraction (36\%) of FCSS objects do not have 2MASS
counterparts. As 2MASS has no bright limit, these must be lost because
they are too faint in the IR (i.e. too blue in optical to
IR colour).

After removing galaxies also present in the XSC we are
left with 228 PSC objects (including 3 QSOs)
which have FCSS redshifts showing
them to be extragalactic ($cz > 900$~km~s$^{-1}$), 
i.e. twice as many galaxies 
as are found in the XSC. As a minimum, 6\% (228/3982) of PSC objects must be extragalactic (11\% of
matched PSC/FCSS objects). This is rather lower than the $\approx$15\% of PSC objects found to be
resolved in SDSS by \citet{finlator00}. Unmatched PSC
objects may include fainter, but still resolved, galaxies; 
almost all the PSC/FCSS objects have $b_j \geq 18.2$ and most
have $18.7 \leq b_j \leq 19.7$.

\begin{figure}
 \resizebox{\hsize}{!}{\includegraphics{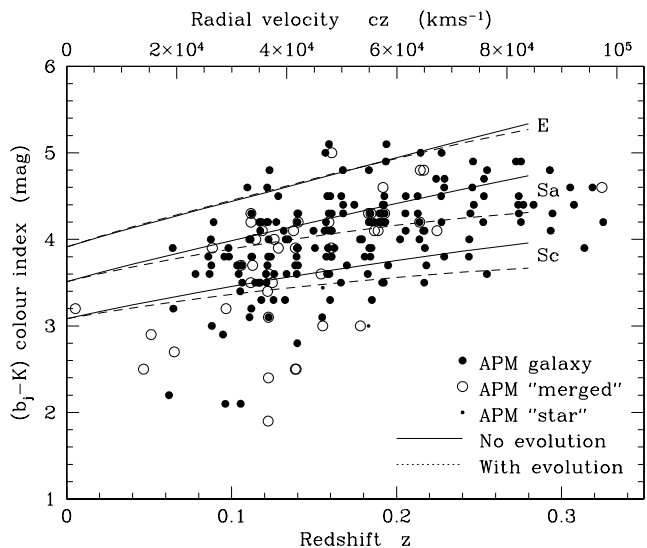}}
 \caption{Optical - IR colour, $b_j-K_s$, versus redshift for
 matched PSC objects with galaxy spectra. Different optical
 image morphologies are indicated. The model tracks are from
 \cite{poggianti97}. }
\label{figure2}
 \end{figure}

The 225 PSC galaxies
(Figure \ref{figure2}) again contain a substantial number classified as
`merged' (36 in the $b_j$ band, though only 13 in $r_f$) and 2 which are optically unresolved (one is an
ultra-compact dwarf (UCD) from \citet{phillipps01} and the other a
compact emission line galaxy from \citet{drinkwater99}). Both merged
and unresolved objects would
typically be ignored in galaxy redshift surveys. The matched sample
has a median $z$ of 0.16, essentially the same as that of the whole FCSS
galaxy sample, and reaches $z = 0.32$.  
However, with many fewer objects over the same redshift
range, the matched sample is obviously picking out only the physically
more luminous, redder, and presumably more massive, galaxies. 2MASS
misses just as large a fraction (or even a larger fraction) of the
nearby FCSS galaxies as distant ones. Indeed, just one cluster
galaxy is detected, the UCD mentioned above.

The PSC sample goes $\sim 1$ magnitude deeper in $K$ than the XSC sample (and 2
magnitudes deeper than the quoted completeness limit of the XSC), though with
less well determined selection criteria. There is rather little overlap in
magnitude with the XSC sample, as the vast majority of the PSC galaxies have
$14.5 \leq K_s \leq 15.5$ (see figure \ref{figure3} below). The numbers as a function
of magnitude suggest that the PSC sample is fairly complete to $K_s \simeq
15.5$. \citet{kochanek01} show that at these magnitudes the canonical 0.6 slope
fits the counts rather well, so going 1 magnitude deeper we
should get a factor 4 more detections, compared to the observed 315/91
= 3.5 for the ratio of XSC and PSC non-cluster galaxies. 
This means that the PSC is a rich source of
moderately faint galaxies not included in the XSC due to
their lack of spatial extension. XSC galaxies represent
only $\sim$ 29\% of the (non-cluster)
galaxies actually measured by 2MASS in this area. 
Virtually all (99\%) of
the galaxies revealed by their spectra could also have been found
purely in terms of optical image
classification  
(provided one
included `merged' images among the potential galaxies). Thus, with
the benefit of the results obtained from all-object surveys such as ours
we extend the search for PSC galaxies by simply matching to,
say, resolved images in APM or SDSS data. Note however, that a substantial
fraction of the PSC galaxies will be fainter than the SDSS or even 2dF
spectroscopic limit, and the `merged' images would remain problematic.

Template matching suggests that in this deeper sample 43\% of the
galaxies (97/225) are now best fitted by a K star
spectrum (slightly fewer than before) and 33\% (75/225) by G or F star
spectra. This implies a significant shift towards the
`intermediate' type spectra for these objects, which are
forced towards slightly bluer $b_j - K_s$ since they are
fainter in $K_s$ than the XSC objects, but have the same $b_j$
limit. Intermediate types predominate amongst the most distant
matched objects; again suggesting the loss of the reddest galaxies.
Only 23\% (53/225) of the PSC/FCSS objects
are fitted by emission line templates
(essentially the same as in the XSC sample), compared to almost 50\%
of FCSS objects as a whole, i.e. 2MASS has a
lower content of star forming galaxies than
does a blue selected sample.
Indeed, 680 of the FCSS galaxies (and 55 AGN) are
{\em not} bright enough in the near IR to enter the 2MASS
catalogues. Thus, only $\sim$ 300 out of 1000 FCSS
galaxies are in 2MASS, whereas $\sim$ 1800 of 2200 stars are
present: in the same, fixed optical magnitude range,
galaxies are far {\em less} likely to be included in the 2MASS
catalogues than stars, again because of a combination of their
surface brightnesses and colours. 
As we have seen, cluster dwarf galaxies are less likely still. 
\footnote{60\% of missed FCSS galaxies 
have emission line spectra, 29\% match G, F or A star
spectra and only 11\% K or M type spectra.}

Given that many galaxies are not large enough to appear in the XSC, it
is appropriate to consider whether unresolved objects have an impact
on the IR galaxy luminosity function. C2001 have
already made careful allowance for this, using the expected isophotal
radii of galaxies to correct for the numbers lost because of the
rather large isophotal diameter limit implied for the XSC
\citep[see also][]{andreon02}. As previously stated, our PSC and XSC samples have little overlap in $K_s$ band
magnitude and no overlap with the brighter, complete sample from the XSC. Thus
few galaxies which should have been present at, say, $K_s=14.5$ have in
actuality been overlooked because of their small size. Thus existing LFs from
data limited at $K_s \leq 14.5$ should not suffer from significant
incompleteness.

The caveat is that there may be galaxies which, because of the
isophotal limit, fail to make the cut for either the XSC or the
PSC. In particular, low surface brightness galaxies could have very
low signal-to-noise ratios in the IR bands. As noted above, our
matched sample actually contains rather few dwarfs, which typically
are of low surface brightness.  As we do not yet have an accurate
distribution of optical to IR colours for dwarfs and other low
surface brightness galaxies, it is not possible to estimate how many
of the optically selected dwarfs should have been included in the
2MASS samples purely on the grounds of their total magnitudes but were
omitted because of their surface brightness. The Millennium Galaxy
Catalogue \citep{driver2005} should give a good indication of the
numbers of low surface brightness objects that might be missed
(assuming plausible optical/IR colours).

\begin{figure}
\resizebox{\hsize}{!}{\includegraphics{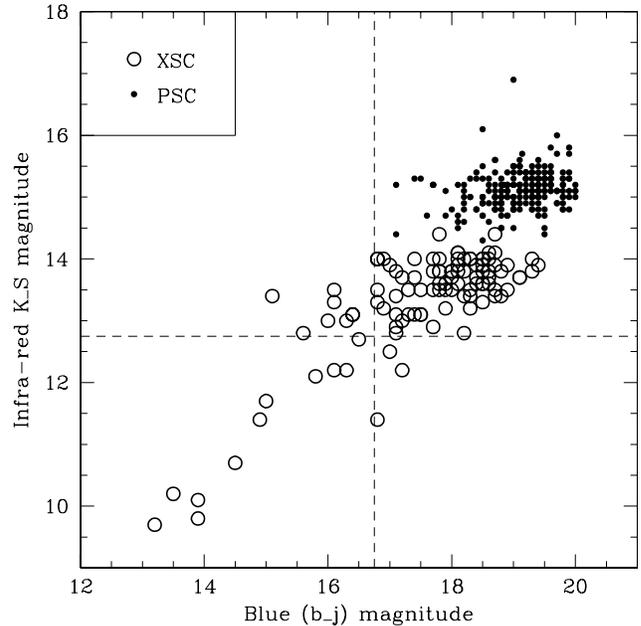}}
\caption{A $b_{j}$ \citep{drinkwater01b} vs $K_{s}$ plot of our XSC
sample, PSC sample and 6dfGRS completeness limits. XSC magnitudes were slightly underestimated in
the 2MASS 2nd data release (see C2001) accounting for some of the gap between
the XSC and PSC galaxies.}
\label{figure3}
\end{figure}

Some implications of this work for other surveys based on the
XSC, for example, the 6dfGRS \citep{jones2004} are shown in figure
\ref{figure3}. The 6dfGRS XSC galaxies were supplemented by other
catalogues including SuperCOSMOS \citep{hambly2001} allowing us to plot
the completeness limits at $b_{j}=16.75$ and
$K_{s}=12.75$. The extra PSC galaxies are below the faint
limit of the 6dfGRS (dashed lines) confirming its completeness to its stated limits.

\begin{acknowledgements}

The FCSS project would not have been possible without the superb 2dF
facility provided by the AAO.MJD thanks the Australian Research
Council for support of this work.MDG acknowledges support from grant
No. 0407445 from the National Science Foundation; part of the work
reported here was done at the Institute of Geophysics and Planetary
Physics, under the auspices of the U.S. Department of Energy by
Lawrence Livermore National Laboratory under contract
No.~W-7405-Eng-48. RAHM and SP thank project students Rebecca Groves
and Yvonne Relf for the initial catalogue matching.

\end{acknowledgements}
\bibliographystyle{aa}
\bibliography{3734}

\end{document}